\documentclass[twocolumn]{aastex631}

\usepackage{amsmath, amsthm, amssymb}
\usepackage{microtype}
\usepackage{float}
\usepackage{verbatim}
\usepackage{graphicx}
\usepackage{multirow}
\usepackage{hhline}
\usepackage{booktabs}
\usepackage{array}
\usepackage{bm}
\usepackage[normalem]{ulem}

\newcommand{\source}{1RXH\,J082623.6$-$505741}

\newcommand{\fglsource}{4FGL\,J0826.1$-$5053}
\newcommand{\accretionratemin}{$3 \times 10^{-11}$}
\newcommand{\accretionratemax}{$3 \times 10^{-10}$}

\renewcommand\ion[2]{#1\,\textsc{\lowercase{#2}}}  

\shorttitle{\source{}: a new evolved cataclysmic variable}
\shortauthors{Sokolovsky et al.}

\begin{document}

\title{\source{}: a new long-period cataclysmic variable with an evolved donor and a low mass transfer rate}

\correspondingauthor{Kirill V. Sokolovsky}
\email{kirx@kirx.net}

\author[0000-0001-5991-6863]{Kirill V. Sokolovsky}
\affiliation{Center for Data Intensive and Time Domain Astronomy, Department of Physics and Astronomy,\\ Michigan State University, East Lansing, MI 48824, USA}
\affiliation{Sternberg Astronomical Institute, Moscow State University, Universitetskii pr. 13, 119992 Moscow, Russia}
\author[0000-0002-1468-9668]{Jay Strader}
\affiliation{Center for Data Intensive and Time Domain Astronomy, Department of Physics and Astronomy,\\ Michigan State University, East Lansing, MI 48824, USA}
\author[0000-0003-1699-8867]{Samuel J. Swihart}
\affiliation{National Research Council Research Associate, National Academy of Sciences, Washington, DC 20001, USA,
resident at Naval Research Laboratory, Washington, DC 20375, USA}
\author[0000-0001-8525-3442]{Elias Aydi}
\affiliation{Center for Data Intensive and Time Domain Astronomy, Department of Physics and Astronomy,\\ Michigan State University, East Lansing, MI 48824, USA}
\author[0000-0003-2506-6041]{Arash Bahramian}
\affil{International Centre for Radio Astronomy Research, Curtin University, GPO Box U1987, Perth, WA 6845, Australia}
\author[0000-0002-8400-3705]{Laura Chomiuk}
\affiliation{Center for Data Intensive and Time Domain Astronomy, Department of Physics and Astronomy,\\ Michigan State University, East Lansing, MI 48824, USA}
\author[0000-0003-3944-6109]{Craig O. Heinke}
\affil{Physics Department, University of Alberta, 4-183 CCIS, Edmonton, AB T6G 2E1, Canada}
\author[0000-0002-1718-0402]{Allison K. Hughes}
\affil{University of Arizona, Department of Physics, 1118 E. Fourth Street, Tucson, AZ 85721, USA}
\author[0000-0001-8229-2024]{Kwan-Lok Li}
\affil{Department of Physics, National Cheng Kung University, 70101 Tainan, Taiwan}
\author[0000-0002-6211-7226]{Raimundo~Lopes de Oliveira}
\affil{Departamento de F\'isica, Universidade Federal de Sergipe, Av. Marechal Rondon, S/N, 49100-000, S\~ao Crist\'ov\~ao, SE, Brazil}
\affil{Observat\'orio Nacional, Rua Gal. Jos\'e Cristino 77, 20921-400, Rio~de~Janeiro, RJ, Brazil}
\author[0000-0003-3124-2814]{James C. A. Miller-Jones}
\affil{International Centre for Radio Astronomy Research, Curtin University, GPO Box U1987, Perth, WA 6845, Australia}
\author[0000-0002-8286-8094]{Koji Mukai}
\affil{CRESST and X-ray Astrophysics Laboratory, NASA/GSFC, Greenbelt, MD 20771, USA}
\author[0000-0003-4102-380X]{David J. Sand}
\affil{Steward Observatory, University of Arizona, 933 North Cherry Avenue, Tucson, AZ 85721-0065, USA}
\author[0000-0003-0286-7858]{Laura Shishkovsky}
\affiliation{Center for Data Intensive and Time Domain Astronomy, Department of Physics and Astronomy,\\ Michigan State University, East Lansing, MI 48824, USA}
\author[0000-0002-4039-6703]{Evangelia Tremou}
\affil{National Radio Astronomy Observatory, P.O. Box O, Socorro, NM 87801, USA}
\author[0000-0001-6215-0950]{Karina Voggel}
\affiliation{Universite de Strasbourg, CNRS, Observatoire astronomique de Strasbourg, UMR 7550, 67000, France}

\begin{abstract}
We report the discovery of \source{}, a 10.4\,hr orbital period compact binary. 
Modeling extensive optical photometry and spectroscopy reveals a $\sim 0.4 M_{\odot}$ K-type secondary transferring mass through 
a low-state accretion disk to a non-magnetic $\sim 0.8 M_{\odot}$ white dwarf. The secondary is overluminous for its mass and dominates 
the optical spectra at all epochs, and must be evolved to fill its Roche Lobe at this orbital period. 
The X-ray luminosity $L_X \sim 1$--$2 \times 10^{32}$ erg s$^{-1}$ derived
from both new XMM-Newton and archival observations,
although high compared to most CVs, still only requires a 
modest accretion rate onto the white dwarf of 
$\dot{M} \sim$\accretionratemin\ to 
\accretionratemax $M_{\odot}$ yr$^{-1}$, 
lower than expected for a cataclysmic variable with an evolved secondary.
No dwarf nova outbursts have yet been observed from the system, consistent with the low derived mass transfer rate.
Several other cataclysmic variables with similar orbital periods also show unexpectedly low mass transfer rates, 
even though selection effects disfavor the discovery of binaries with these properties. 
This suggests the abundance and evolutionary state of long-period, low mass transfer rate cataclysmic variables is worthy of additional attention.
\end{abstract}

\section{Introduction}
\label{sec:intro}

\subsection{Observational appearance of cataclysmic variables}
\label{sec:cvobs}

Binary stars containing a white dwarf accreting from 
a dwarf of a subgiant companion are called cataclysmic variables 
\citep{1995cvs..book.....W,2001cvs..book.....H}.
They earned this name thanks to two types of violent phenomena that may dramatically 
increase optical brightness of such binaries: classical nova eruptions and dwarf nova outbursts. 
A classical nova eruption occurs when thermonuclear reactions ignite in the accreted 
hydrogen-rich shell \citep{2008clno.book.....B,2016PASP..128e1001S,2021ARA&A..59..391C},
typically leading to a high ($\sim 8$--15 mag; \citealt{1990ApJ...356..609V,2008clno.book.....W,2021ApJ...910..120K}) outburst amplitude.
A dwarf nova outburst is powered by a completely different physical mechanism.
These more modest outbursts (typically $\sim 2$--8 mag; 
\citealt{2016MNRAS.456.4441C,2021ApJ...910..120K})
occur when the accretion disk surrounding a white dwarf switches from 
a low-viscosity, low-accretion-rate state with an emission-line dominated
spectrum\footnote{The emission line spectrum does not necessary mean  
that the whole disk is transparent. It implies that there must 
be a temperature inversion somewhere in the disk atmosphere. 
In the standard disk instability model the quiescent dwarf nova disk 
is opaque \citep{2010A&A...519A.117I}. 
However, the observations suggest it is at least partly
transparent
\citep{1999ApJ...523..399F,2001MNRAS.327..475L,2007A&A...475..301K,2010MNRAS.402.1824C}.} 
to a high-viscosity, high-accretion-rate continuum-dominated state. 
This disk instability model for the dwarf
nova phenomenon has been extensively discussed in the literature \citep{1996PASP..108...39O,2005PJAB...81..291O,2020AdSpR..66.1004H,2007A&ARv..15....1D}.
Dwarf novae occur if the mass transfer rate is insufficient to maintain 
the accretion disk permanently in the hot high-viscosity state and if the accretion disk formation 
is not disrupted by the strong magnetic field of the white dwarf
\citep{1990SSRv...54..195C,1994PASP..106..209P,2015SSRv..191..111F}.

Cataclysmic variables are often prominent X-ray sources \citep{2017PASP..129f2001M}. 
While in classical novae X-rays are emitted by plasma heated by nuclear burning and shocks within the nova ejecta, 
X-rays from non-nova cataclysmic variables are powered by accretion.
Specifically, the X-rays emerge at the interface between 
the accretion disk/stream and the white dwarf surface, known as the boundary layer. 
The exact structure of this region, the role of shocks and magnetic field as well as changes
accompanying the disk state transitions are debated 
\citep[e.g.,][]{1995ApJ...442..337P,2005ApJ...626..396P,2010SSRv..157..155I,2021ApJ...918...87D}.
The optically thin thermal X-ray emission of cataclysmic variables 
has the maximum temperature k$T_{\rm max}=$6--55\,keV 
and luminosity $10^{28}$--$10^{32}$\,erg\,s$^{-1}$ \citep[][]{2020AdSpR..66.1097B}. 
If the accretion disk is disrupted by the white dwarf's magnetosphere, 
the matter reaches the white dwarf surface via the accretion columns above 
the magnetic poles \citep{2000SSRv...93..611W,2012MmSAI..83..578M}. 
The X-ray emission of magnetic cataclysmic variables is characterized by a maximum temperature of a few tens of keV 
and spans the range of luminosities from $10^{30}$ to $10^{34}$\,erg\,s$^{-1}$ \citep{2020AdSpR..66.1209D}.
The white dwarf rotation moves the accretion columns in and out of view, 
modulating the X-ray flux. These periodic variations, together with the generally 
harder X-ray spectra, are the telltale signs of magnetic cataclysmic variables.

\subsection{Origin and evolution of cataclysmic variables}

Cataclysmic variables are thought to have 
evolved from binaries that survived a common envelope 
phase without merging \citep{1976IAUS...73...75P,2010MmSAI..81..849R,2013A&ARv..21...59I}. 
The typical stages are as follows. As the more massive star in a wide 
separation binary evolves into a red giant, it may overflow its Roche lobe,
starting mass transfer to its less massive companion. As the convective
envelope of the red giant expands due to mass loss on a dynamical timescale
and the binary orbit shrinks
due to conservation of angular momentum, this is a runaway process \citep{1972AcA....22...73P}, causing the
components to be engulfed in a common envelope on a dynamical timescale. The successful ejection
of the envelope will lead to a hardened binary containing a white dwarf
and the original secondary.

While mass transfer from more massive to less massive binary component is 
unstable, the opposite process is sustainable, and in principle
mass transfer can begin from the secondary to the white dwarf if the secondary
fills its Roche Lobe during main sequence or later evolution.
This is the stage where the binary would become observable as a cataclysmic variable.
Transferring mass from a less massive to more massive component brings 
the accreted matter closer to the binary center of mass, requiring 
the component separation to increase in order to conserve angular momentum, 
which tends to break contact between the binary components. An additional physical 
mechanism, either angular momentum loss or an expansion of the secondary in response to mass loss,
is needed to restore and maintain contact.

\subsubsection{The canonical evolution path}

The common wisdom is that cataclysmic variable evolution is driven by secular angular momentum loss 
that keeps the binary components in contact \citep[e.g.,][]{2011ApJS..194...28K}. 
The two relevant physical mechanisms 
are gravitational wave emission \citep{1967AcA....17..287P,1981ApJ...248L..27P} 
dominating in systems with  orbital periods less than 3~hours, and 
magnetic braking \citep{1968MNRAS.138..359M,1981A&A...100L...7V} dominating at longer periods. 
Magnetic braking occurs due to magnetic coupling between the star and 
its ionized stellar wind carrying away angular momentum. As the secondary star in 
a cataclysmic variable is tidally locked (its axial rotation period is equal 
to the orbital period), the angular momentum loss due to the magnetized wind 
has to be compensated by decreasing orbital period. 
Once the secondary star loses enough mass to become fully 
convective (corresponding to the orbital period of about 3 hours), 
the effectiveness of the magnetic braking appears to be dramatically 
reduced \citep{1983A&A...124..267S,1984MNRAS.209..227V}.
Most interacting systems lose contact when evolving to periods shorter than 3~hours
and remain detached until gravitational wave emission brings 
the components back in contact at periods of about 2~hours,
leading to an observed gap in cataclysmic variable periods between $\sim 2$ and 3 hours \citep{1998MNRAS.298L..29K,2001ApJ...550..897H}.

It is possible that other angular momentum loss mechanisms 
operate in cataclysmic variables in addition to  gravitational radiation 
and magnetic braking. Angular momentum loss may happen during nova eruptions \citep{1998MNRAS.297..633S,2016MNRAS.455L..16S,2022MNRAS.510.6110P}. 
Indeed, each nova eruption is akin to a common envelope phase \citep{2016ApJ...817...69N,2021ApJ...914....5S}, 
with the white dwarf atmosphere expanding to engulf both binary components.
Other possibilities discussed in the literature include angular momentum 
losses via a circumbinary disk \citep{2005ApJ...635.1263W} and magnetized 
accretion disk wind \citep{1988ApJ...327..840C}.
Nova eruptions may also heat (and inflate) the secondary, increasing the mass 
transfer rate above its mean long-term value \citep{2021MNRAS.507..475G}. 

Cataclysmic variables evolve towards short periods until the thermal timescale of the secondary
becomes longer than the mass transfer timescale, which leads to an increase rather than a decrease in the orbital period 
in response to mass loss. The minimum 
period is about 80 minutes \citep[e.g.,][]{1985SvAL...11...52T,2009MNRAS.397.2170G} for 
the ordinary hydrogen-rich donors and is shorter for helium-rich donors 
(that have smaller sizes for a given mass).

\subsubsection{The alternative: nuclear evolution of the donor}

The contact between the binary components can be maintained 
without major angular momentum loss if the donor is expanding due to nuclear evolution.
A well-understood case is if the donor is less massive than the white dwarf and can 
transfer mass as a subgiant or red giant, with the orbital period and mass transfer
rate growing with time \citep{1983ApJ...270..678W,1988A&A...191...57P,2015ApJ...809...80G}.
Another possibility arises if the donor star is more massive than the white
dwarf after the common envelope phase. In this case, there can be an initial phase
of thermal timescale mass transfer once the donor fills its Roche Lobe as a subgiant. 
This reduces the orbital period until the two components have similar masses. 
After that, mass transfer proceeds on a timescale governed by magnetic braking and/or nuclear evolution 
(e.g., \citealt{2004ApJ...601.1058I,2016ApJ...833...83K}).

\subsection{\source{} and the scope of this work}

\source{} originally drew our attention as 
a possible X-ray counterpart of the Fermi-LAT $\gamma$-ray source 3FGL\,J0826.3$-$5056. 
Once our initial optical spectroscopy showed it to be a Galactic source with evidence for ongoing accretion, we started a multiwavelength campaign. 
With the subsequent release of the Fermi-LAT Pass~8 data \citep{2018arXiv181011394B}, and eventually 
the 4FGL (and 4FGL Data Release~2) catalogs \citep{2020ApJS..247...33A,2020arXiv200511208B}, the location of 
the $\gamma$-ray source ``moved'' to the northeast, leaving \source{} well outside the 95\% error ellipse of 
the revised $\gamma$-ray position (now dubbed \fglsource{}).
%
The optical Gaia~EDR3 \citep{2020arXiv201201533G} counterpart of \source{} has an ICRS position of 
08:26:23.801, $-$50:57:43.46
at the mean epoch 2016.0 and a highly significant parallax measurement ($\varpi = 0.583\pm0.036$\,mas), 
resulting in a distance of $1.63\pm0.09$\,kpc after applying a zeropoint offset of $-0.0294$\,mas \citep{2020arXiv201201742L}; 
because of the precise measurement the prior on the distance is not important.
The mean optical magnitude of \source{} is $V = 16.60 \pm 0.06$ 
\citep[][]{2015AAS...22533616H}, 
corresponding to an absolute magnitude of $M_V \sim 5.0$--5.2, depending on the extinction adopted (see discussion in Section~\ref{sec:extinct}).  

Here we report on optical, X-ray and radio observations of \source{} that, although unrelated to \fglsource{}, 
was revealed to be a cataclysmic variable star with an evolved secondary. 
We derive the parameters of the binary system and discuss them in 
the context of cataclysmic variable population and evolution.
The inferred low mass transfer rate of this binary is challenging to explain with 
current cataclysmic variable evolution theory.

\section{Observations}

\subsection{X-ray}

\subsubsection{XMM-Newton}
\label{sec:xmmobs}

We performed a dedicated XMM-Newton observation of \source{} (as a candidate GeV source counterpart;
Section~\ref{sec:intro}) for 48\,ks starting on 
 2017-06-13 21:44:37~TDB (ObsID~0795710101). The observation start time  corresponds to the orbital phase of 0.76 and covers almost 1.3 orbital periods of the system
(Section \ref{sec:rv}). The Medium filter was used for the
three European Photon Imaging Camera (EPIC) instruments. 
The source is clearly visible with 
1437 ($0.021 \pm 0.001$\,cts/s), 1527 ($0.025 \pm 0.001$\,cts/s) and 3404
($0.083 \pm 0.003$\,cts/s) photons attributable to the source detected by the first and second 
EPIC-Metal Oxide Semiconductor (MOS1 and MOS2; \citealt{2001A&A...365L..27T}) and 
EPIC-pn \citep{2001A&A...365L..18S} cameras, respectively, in the 0.2--10\,keV energy range. 
The source is too faint for grating spectroscopy with the Reflection Grating Spectrometers \citep{2001A&A...365L...7D}.

We used \textsc{xmmsas\_20190531\_1155-18.0.0} \citep{2004ASPC..314..759G} for reducing the EPIC data.
After the initial data ingestion \textsc{cifbuild}, \textsc{odfingest}, and pipeline-processing \textsc{emproc}/\textsc{epproc} steps, 
we constructed pattern zero (single-pixel events) only, high energy 
lightcurves used to identify times of high particle background. 
To construct these lightcurves, we apply filtering on the Pulse Invariant channel number (\texttt{PI}; encoding the photon energy) 
and pixel pattern triggered by a photon or a background charged particle (\texttt{PATTERN}; sometimes referred to as ``event grade''). We apply the following \textsc{evselect} filtering expressions: \texttt{(PI>10000) \&\& (PATTERN==0)}
for MOS1/2 and \texttt{(PI>10000 \&\& PI<12000) \&\& (PATTERN==0)} for pn cameras, respectively.
After a visual inspection, we selected low background good time intervals
with \texttt{RATE<=0.2}, \texttt{<=0.25}, \texttt{<=0.4} for the three cameras, respectively.
Comparing these thresholds to reference values in the EPIC analysis 
thread\footnote{\url{https://www.cosmos.esa.int/web/xmm-newton/sas-thread-epic-filterbackground}},
the MOS1/2 cut-off levels are relatively aggressive, while the pn cut-off
matches the reference value. These cut-offs result in about 5\% of MOS1/2
and almost 50\% of pn data being rejected.

We converted the photon arrival times to the Solar system barycenter with \textsc{barycen}.
After producing the images with \textsc{evselect} we used \textsc{ds9} to
center a 20\arcsec\ radius circular aperture on the source image
(independently for the three instruments). For background extraction, we used
a source-centered annulus with the inner and outer radii of 39\arcsec\ and
91\arcsec\ (140\arcsec\  and 162.5\arcsec)
for MOS1 (MOS2) and an off-source circle 94\arcsec\ radius circle for pn.
These regions were chosen to avoid nearby sources and check that the
analysis results do not depend strongly on the specific choice of the background
region. 
We extracted source and background spectra from the respective regions with \textsc{evselect} 
applying filtering: \texttt{(PI>150)} for all three cameras; \texttt{(PATTERN<=12)} for MOS1/2 and
\texttt{(PATTERN<=4)} for pn cameras, respectively.
The associated calibration parameters were generated with 
\textsc{backscale}, \textsc{rmfgen} and \textsc{arfgen}.
We used \textsc{specgroup} to bin the spectra to contain at least 25 background-subtracted counts per bin
and not to oversample the intrinsic energy resolution by a factor $>3$.
We also extracted event files (unbinned lightcurves) for the source and background regions.
Finally, we used \textsc{evselect} (\texttt{(PATTERN<=4) \&\& (PI in [200:10000])})
and \textsc{epiclccorr}
to construct background-subtracted 0.2--10\,keV lightcurves with 300\,s time binning
(so on average, the lightcurves contain 6.3, 7.5 and 24.9 counts per bin for MOS1/2, pn). 

\begin{figure}
    \includegraphics[width=1.0\linewidth]{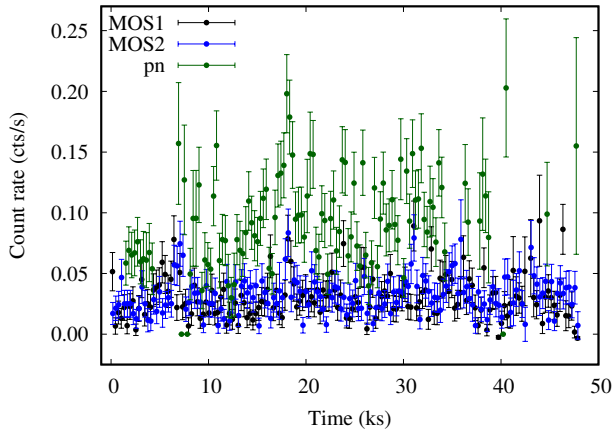}
    \caption{XMM-Newton/EPIC 0.2--10\,keV background-subtracted lightcurve of \source{}. A minimal amount of the MOS 1/2 data was rejected due to background contamination, but the figure is much larger for the pn data, nearly 50\%. Nonetheless, the vast majority of the potential 300-s bins for the pn light curve are usable, as they had at least a partial uncontaminated time interval. There is no systematic change in the count rate during the observation. }
    \label{fig:epiclc}
\end{figure}

Figure~\ref{fig:epiclc} presents the binned lightcurve (48\,ks) of \source{}. 
No systematic change in brightness is seen during the observation, 
however the $\chi^2$ test indicates that the lightcurve scatter is significantly
larger than expected from the error bars
\citep{2010AJ....139.1269D,2017MNRAS.464..274S}. 
This suggests the presence of irregular variations on timescales
comparable or longer than the 300\,s bin size and shorter than the 48\,ks duration of the lightcurve.

The discrete Fourier transforms of the lightcurves \citep{1975Ap&SS..36..137D,2018ApJS..236...16V} 
show no obvious periodicities in the range 600\,s ($2\times$time bin width) to 5000\,s
(1/10 of the observation duration). The need to obtain a significant
estimate of the count rate within a bin limits our ability to search for
periods about or shorter than the bin width. This can be avoided by
performing the period search on the photon arrival times recorded in the event file
directly, with no binning \citep[e.g.,][]{2018JGRA..123.9204J}. 

Background subtraction is not possible for the unbinned lightcurve 
as it is not clear if a given event recorded within the source aperture 
should be attributed to the source or background. This is not a show stopper 
as the constant background should just reduce the pulsed fraction and add
noise to the periodic signal. The background flares (produced by an increased number of charged particles hitting the detector) are not periodic, but 
may introduce ``red-noise'' in the power spectrum with some of its power leaking to high frequencies where we conduct the period search
(see the discussion in Appendix~\ref{sec:hm}). While the time intervals obviously affected by an increased particle background are rejected from the analysis, some residual particle background variations remain. Period searches on the events extracted from the background region are used to identify any spurious periodicities introduced by residual background variations and the efforts to reject visible background flares, as well as check for the presence of any periodic
instrumental effects. 

We performed a discrete Fourier transform (Rayleigh test) 
and an ``$H_m$-periodogram'' search in the trial period range from 0.5\,s
(3.0\,s for MOS1/2 as ``Full Frame'' mode frame time is 2.7\,s) to 5000\,s as described in Appendix~\ref{sec:hm}.
No periodogram peak was be identified that was both significant 
according to the $H_m$ statistic \citep{1989A&A...221..180D,2010A&A...517L...9D,2011ApJ...732...38K} 
and appeared at similar frequencies in lightcurves collected with multiple instruments. 
We confirmed this conclusion by repeating the analysis with another code,
implementing the closely related ``$Z_m^2$ periodogram'' \citep[][see the discussion in Appendix~\ref{sec:hm}]{1983A&A...128..245B} 
applied to the pn data in the 0.3--8\,keV, 0.3--2\,keV, and 2--8\,keV energy ranges. 

Previously, five XMM-Newton slews \citep{2008A&A...480..611S} over the \source{} area in 2005--2013 lead to
$2\sigma$ 0.2--12\,keV EPIC-pn upper limits of 
$\lesssim 1$\,cts/s, according to the Upper Limit Server\footnote{\url{http://xmmuls.esac.esa.int/hiligt/}} \citep{2011ASPC..442..567S}.
Therefore, the source could not have been much brighter during these slews
compared to the long XMM-Newton observation.

\begin{figure}
    \includegraphics[height=1.0\linewidth,clip=true,trim=0cm 0cm 0cm 0cm,angle=270]{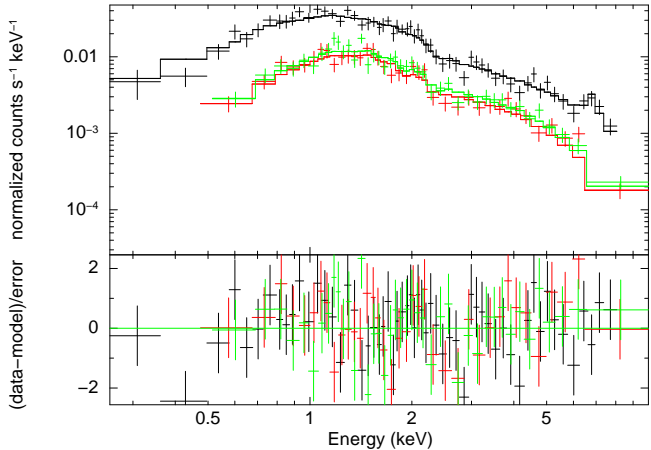}
    \caption{XMM-Newton spectra of \source{} obtained with the
three EPIC cameras (pn -- black, MOS1 -- red, MOS2 -- green) compared to the
absorbed optically thin thermal emission model.}
    \label{fig:epicspectra}
\end{figure}

The spectra obtained with the three EPIC cameras during the dedicated
observation on 2017-06-13 are grouped with \textsc{specgroup} to contain at least 25 counts per energy bin. 
We adopt $\chi^2$ as the fit and test statistic and fix the elemental abundances to the solar values as derived by \cite{2009ARA&A..47..481A}.
We simultaneously fit the spectra obtained by the three EPIC cameras 
with an absorbed single-temperature optically thin thermal emission model (Figure~\ref{fig:epicspectra}). 
Specifically, we use the \textsc{XSPEC} model \texttt{const*phabs*cflux(apec)} where
\texttt{const} is an offset between the cameras, 
\texttt{phabs} is the photoelectric absorption model \citep{1992ApJ...400..699B},
\texttt{cflux} is the convolution model needed only to calculate the flux of 
the next model component, and \texttt{apec} is the collisional ionization equilibrium
plasma emission model that includes both bremsstrahlung continuum and line emission \citep{2005AIPC..774..405B}. 
The fit results in $\chi^2/N_{\rm d.o.f.}=122.4/133$). 
%
The best-fit model parameters and their associated $1\sigma$ uncertainties are: 
the equivalent hydrogen absorbing column density 
$N_\mathrm{H} = 3.2_{-0.3}^{+0.3} \times 10^{21}$\,cm$^{-2}$, 
plasma temperature ${\rm k}T = 9.9_{-1.9}^{+3.5}$\,keV, 
unabsorbed flux $4.9_{-0.2}^{+0.2} \times 10^{-13}$\,erg\,cm$^{-2}$\,s$^{-1}$ at 0.2--10\,keV.
The scaling factor was fixed to 1.0 for the pn camera and the best-fit values
were 0.95 and 1.04 for MOS1 and MOS2.
The Fe~K$\alpha$ emission at 6.7\,keV is clearly visible in the pn spectrum (Figure~\ref{fig:epicspectra})
and is well described by the \texttt{apec} model with \cite{2009ARA&A..47..481A} abundances.

The fitted $N_\mathrm{H}$ is consistent within one sigma with the integrated Galactic
$N_\mathrm{HI}$ value derived from 21\,cm hydrogen line observations  
\citep{2005A&A...440..775K,2005A&A...440..767B,2015A&A...578A..78K}. 
However, much of this absorbing material likely lies beyond the binary. 
Instead, using the $E(B-V)$ value favored from the spectroscopy
(Section~\ref{sec:extinct}) and the relation of \cite{2009MNRAS.400.2050G} between 
the X-ray absorbing column and optical extinction ($N_\mathrm{H} = 2.21 \times 10^{21}\,{\rm cm}^{-2}  A_V$), 
we infer a foreground $N_\mathrm{H} \sim 7 \times 10^{20}$--$1.2\times10^{21}$ cm$^{-2}$, 
depending on which interstellar \ion{Na}{I} absorption to dust extinction
relation we adopt (see Section~\ref{sec:extinct}). \cite{2009MNRAS.400.2050G}
report a four per cent uncertainty in the slope of the $N_\mathrm{H}(A_V)$
relation. It is unclear what is the intrinsic scatter of this relation as 
the individual pairs of $N_\mathrm{H}$ and $A_V$ measurements used by \cite{2009MNRAS.400.2050G} 
have large uncertainties. 
\source{} is located in a scenic region of the sky next to a dark nebula                                                             
and multiple Herbig-Haro objects including HH~46/47 
\citep[6$\arcmin$ away in the foreground at 450\,pc][]{2004ApJS..154..352N}.
The extinction law and dust to gas ratio variations found in 
the vicinity of star forming regions \citep[e.g.,]{2015MNRAS.449.3867G,2015A&A...578A.131L,2017MNRAS.471L..52T} 
further complicate the optical extinction.
While the overall uncertainty of 
the $A_V$-based estimate of $N_\mathrm{H}$ is difficult to
quantify, it is unclear if it can account for almost a factor of three difference 
with the X-ray-derived $N_\mathrm{H}$. 
Therefore, a fraction of the measured $N_\mathrm{H}$ may be intrinsic to \source{}.

At the Gaia parallax distance to the source, the inferred unabsorbed X-ray flux corresponds to 
a 0.2--10\,keV luminosity of $L_X = (1.6\pm0.1) \times10^{32}$ erg s$^{-1}$, which we use for the remainder of the paper.

As the X-ray spectra of many cataclysmic variables are described in terms of
the multi-temperature the cooling flow model \citep[e.g.,][]{2020AdSpR..66.1097B}, 
we fit the model \texttt{constant*phabs*mkcflow} \citep{1988ASIC..229...53M} to the EPIC data.
The highest temperature of the cooling flow is k$T = 43 \pm 10$\,keV 
and the absorbing hydrogen column $N_\mathrm{H} = 3.6_{-0.3}^{+0.3} \times 10^{21}$\,cm$^{-2}$
(close to the one inferred from the single-temperature model).
The accretion rate derived from \texttt{mkcflow} is $3\times10^{-11} M_{\odot}$ yr$^{-1}$, 
about the same as what we derive from the \texttt{apec} fit.
The fit is acceptable with $\chi^2/N_{\rm d.o.f.}=86.46/107$, however, it does not improve 
on the simpler single-temperature plasma model that we use throughout this paper.

\subsubsection{Swift}
\label{sec:swiftobs}

\begin{figure}
    \includegraphics[height=1.0\linewidth,clip=true,trim=0cm 0cm 0cm 0cm,angle=270]{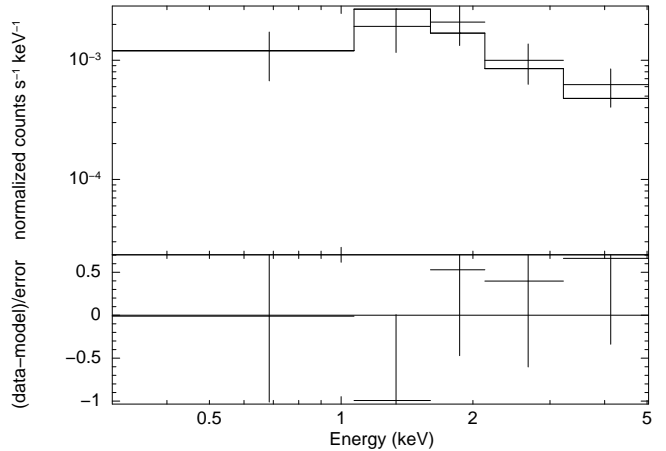}
    \caption{The stacked Swift/XRT spectrum of \source{} compared to the absorbed optically thin thermal emission model.}
    \label{fig:srtspectrum}
\end{figure}

The Neil Gehrels Swift Observatory \citep{2004ApJ...611.1005G} 
observed \source{} in the Photon Counting mode for a total of 7.5\,ks 
between 2015-03-25 and 2017-03-03 (Table~\ref{tab:swiftlog}). 
The observations were obtained in the framework of a program to 
follow-up Fermi unassociated sources\footnote{\url{https://www.swift.psu.edu/unassociated/}} \citep{2013ApJS..207...28S}.
\source{} is clearly visible in the stacked Swift
X-ray Telescope \citep[XRT;][]{2005SSRv..120..165B} 0.3-10\,keV 
image with the net count rate of $0.0071 \pm 0.0012$\,cts/s.
We used \textsc{WebPIMMS}\footnote{\url{https://heasarc.gsfc.nasa.gov/cgi-bin/Tools/w3pimms/w3pimms.pl}}
to predict the XMM-Newton/EPIC-pn $+$ Medium filter $+$ Full Frame mode count rate of
0.081 assuming the thermal plasma model from Section \ref{sec:xmmobs}.
This closely matches what was actually observed by XMM-Newton/EPIC-pn
(Section~\ref{sec:xmmobs}), suggesting a similar X-ray flux in the Swift/XRT data compared to the XMM-Newton data.
We group the stacked Swift/XRT spectrum with \textsc{grppha} to contain at least 9 counts per bin and fit with the
model used in Section \ref{sec:xmmobs} for the XMM-Newton observations
fixing the ${\rm k}T$ and $N_\mathrm{H}$ values and leaving only the model
normalization factor free to vary 
($\chi^2/N_{\rm d.o.f.}=1.86/4$; 
Figure~\ref{fig:srtspectrum}), indicating no compelling evidence for a different spectrum in the Swift/XRT data compared to the higher signal-to-noise XMM data.

\begin{table}
\centering
\caption{Swift observing log}
\begin{tabular}{ccc}
\hline
Obs.ID   & Date & Exposure (s) \\
\hline
\hline
00084684001 & 2015-03-25 &  763  \\
00084684002 & 2015-06-01 &  393  \\
00084684003 & 2015-08-30 &  183  \\
00084684004 & 2015-09-23 &  597  \\
00084684005 & 2015-10-01 & 1660  \\
00084684006 & 2017-03-03 & 3939  \\
\hline
\end{tabular}
\label{tab:swiftlog}
\end{table}

\subsubsection{ROSAT}
\label{sec:rosatobs}


The Upper Limit Server 
lists a pointed 20\,ks ROSAT/HRI observation of the \source{} area on 1994-05-12.
The source is detected with the 0.2--2\,keV HRI count rate of 
$0.0012 \pm0.0004$\,cts/s. 
The source is listed as \source{} by \cite{2000yCat.9028....0R}. 
The upper limits server lists no pointed ROSAT/PSPC data,
however the source was observed in the survey mode
\citep{1999A&A...349..389V,2000IAUC.7432....3V,2016A&A...588A.103B} on 1990-10-30 
resulting in a $2\sigma$ 0.2--2\,keV PSPC upper limit of $<0.022$\,cts/s.
Using \textsc{WebPIMMS} 
to extrapolate the ROSAT/HRI detection to the XMM-Newton/EPIC-pn band
using the absorbed thermal plasma model derived from the XMM-Newton
observations described in Section \ref{sec:xmmobs}, we estimate 
XMM-Newton/EPIC-pn $+$ Medium filter $+$ Full Frame mode count rate of
0.038\,cts/s, a factor of two lower than what was actually observed.
This may suggest factor of $\sim 2$ X-ray flux variability 
on a decades-long timescale.

\subsection{Optical Spectroscopy}
\label{sec:optobsspec}

\subsubsection{SOAR}

\begin{figure}
    \includegraphics[width=1.0\linewidth]{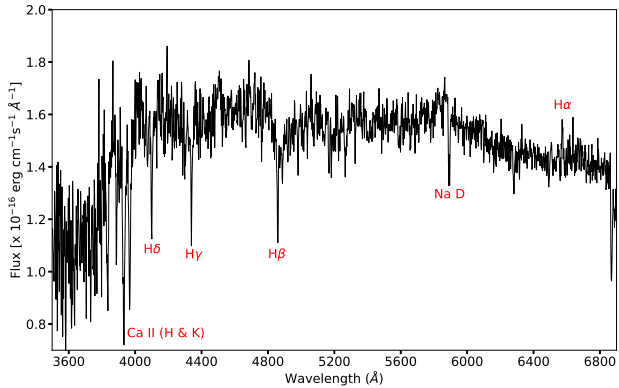}
    \caption{Low-resolution SOAR/Goodman spectrum of \source{} from 2016-12-30. At this epoch the emission from the accretion disk was relatively faint compared to the cool secondary star, which dominates the optical emission. An approximate flux calibration has been applied to the data.}
    \label{fig:soarspec}
\end{figure}

Using the Goodman Spectrograph \citep{2004SPIE.5492..331C} mounted on the
4.1\,m SOAR telescope, we obtained optical spectroscopy of \source{} on parts of 
13 nights from 2016-12-30 to 2019-11-06. The main setup used a 1200\,l\,mm$^{-1}$ grating with 
a 0.95\arcsec\ slit, giving a full-width at half-maximum resolution of about
1.9\,\AA\ over the wavelength range 5500--6750\,\AA. 
Depending on the seeing, the exposure time per spectrum was 900 or 1200\,s. 
Spectral reduction and optimal extraction were performed in the usual manner
using {\tt IRAF} \citep{1986SPIE..627..733T}.
A SOAR/Goodman spectrum covering a wider wavelength range of 3600--6850\,\AA\ (400 l\ mm$^{-1}$ grating, 1\arcsec\ slit) obtained on
2016-12-30 is
presented in Figure~\ref{fig:soarspec}.

\subsubsection{Magellan}
\label{sec:magellan}

To assess the rotational broadening of the secondary, we obtained on 2019-04-06 a single 2400\,s high-resolution spectrum of 
the system with MIKE \citep{2003SPIE.4841.1694B} on the Magellan/Clay Telescope. Data were accumulated on both the blue and red sides of the instrument, 
with resolutions of $R \sim 30000$ and $\sim 26000$, respectively. The data were reduced using {\tt CarPy} \citep{2003PASP..115..688K}.

\subsection{Optical Photometry}
\label{sec:optobsphot}

\begin{figure}
    \includegraphics[width=1.0\linewidth]{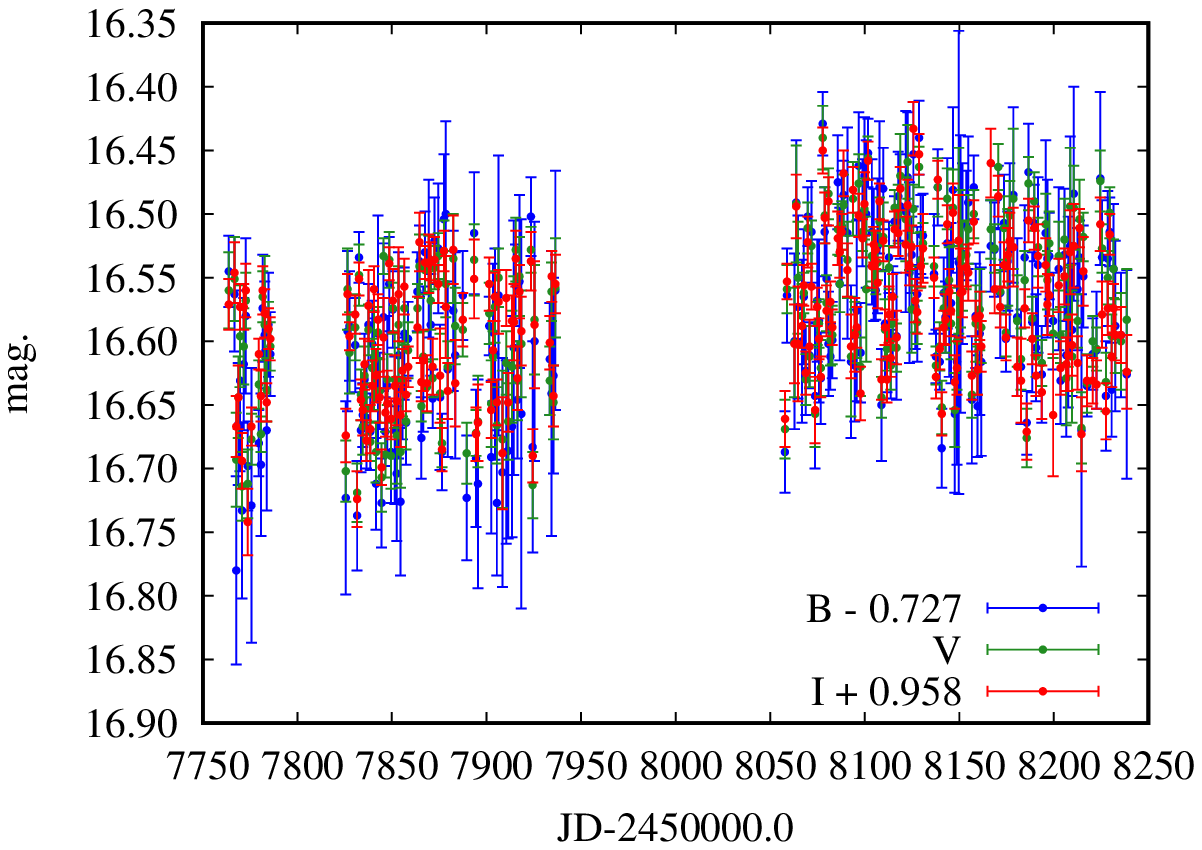}
    \caption{SMARTS 1.3\,m ANDICAM $BVI$ lightcurve of \source{}.}
    \label{fig:opttimelc}
\end{figure}

We obtained 226 $B$, 228 $V$ and 226 $I$ band photometric measurements 
using ANDICAM on the SMARTS 1.3\,m telescope at CTIO 
over 228 nights between 2017-01-08 and 2018-04-29. 
The data reduction is described by \cite{2012PASP..124.1057W}.
The estimated photometric uncertainties are 0.050, 0.026 and 0.022\,mag in $B$, $V$ and $I$ bands, respectively.
The lightcurve (Figure~\ref{fig:opttimelc}) shows a $0.06$\,mag increase in mean brightness over 
the duration of the observations as well as a short-term modulation, but no flaring activity.

\begin{figure}
    \includegraphics[width=1.0\linewidth]{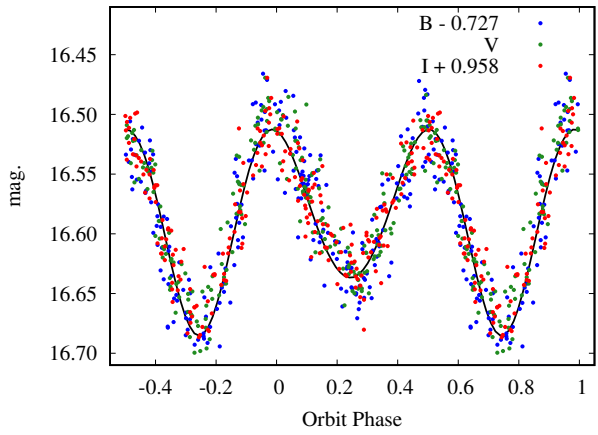}
    \caption{SMARTS 1.3\,m ANDICAM $BVI$ lightcurve of \source{} phased with the
photometric period (\ref{eq:lightents}) and the spectroscopically derived
minimum radial velocity phase $T_{0~{\rm BJD(TDB)}}$ (Section~\ref{sec:rv}) in  for consistency with
Figure~\ref{fig:opt_spectra}. The curve represents the model described in
Section~\ref{sec:binarymodel}.}
    \label{fig:optphaselc}
\end{figure}

After detrending the lightcurve with a piecewise linear function,
applying the barycentric correction \citep[e.g.,][]{2010PASP..122..935E} 
and combining $B$, $V$ and $I$ lightcurves removing the median offsets of 
$(B-V) = 0.727$ and $(V-I) = 0.958$, 
we performed a period search using the string-length technique of
\cite{1965ApJS...11..216L} resulting in the following 
light elements:
\begin{equation}
\label{eq:lightents}
{\rm BJD}_{\rm min}{\rm (TDB)} = 2457831.62700  + (0.431698 \pm 0.000020) \times {\rm E}
\end{equation}
We identify the light variations period with the orbital period of the binary
system, the conclusion confirmed with spectroscopy discussed in Section~\ref{sec:rv}.
The period uncertainty is calculated allowing for the maximum phase shift of
0.05 between the first and the last observation, see Appendix~\ref{sec:hm}.
The phased lightcurve is presented in Figure~\ref{fig:optphaselc}.
The epoch of primary minimum in (\ref{eq:lightents}) corresponds to 
the orbit phase of 0.75 defined in Section~\ref{sec:rv}. 

We choose the \cite{1965ApJS...11..216L} period search technique as it is
suitable for phased lightcurves of an arbitrary form (they do not have to be
sine-wave like) and the lightcurve at Figure~\ref{fig:optphaselc} has unequal
minima. While the visual inspection of the phased lightcurve leaves no doubt
that the brightness variations are real, we confirm this by performing the formal tests.
We use bootstrapping \citep[e.g., \S~7.4.2.3. of][]{2018ApJS..236...16V} 
to test the chance occurrence probability of the periodicity detected
with the \cite{1965ApJS...11..216L} technique. We also construct 
the Lomb-Scargle periodogram \citep{Lomb_1976,1982ApJ...263..835S,2018ApJS..236...16V}
and use eq.~(14) of \cite{1982ApJ...263..835S} to estimate the significance of
its highest peak (corresponding to half the orbital period).
We follow Appendix~\ref{sec:hm} to estimate the number of independent
frequencies needed for the Lomb-Scargle peak significance calculation.
Both bootstrapping and the analytical calculation point to the chance
occurrence probability of the periodic signal to be $\ll10^{-4}$.

Nothing is visible in the GALEX near-UV (2310\,\AA) image covering the source position
\citep{2017ApJS..230...24B}. The XMM Optical Monitor observations that accompanied our X-ray data were taken in $V$ rather than in the UV
and hence offer no addition UV constraints.

\subsection{Radio}
\label{sec:radioobs}

We obtained radio continuum observations of \source{} with the Australia Telescope Compact Array
on 2017-03-03, while the array was in its 6D configuration 
(5.9\,km maximum antenna spacing), and using the 5.5/9.0 GHz backend (we only consider the lower-noise 5.5 GHz observations here). The on-source integration time was 10.2 hr, and we used 
PKS\,0823$-$500 as the complex gain calibrator and
PKS\,1934$-$638 as the flux density and bandpass calibrator. The data were reduced and imaged using standard tasks in
\textsc{CASA} \citep{2007ASPC..376..127M}, with the imaging done with a Briggs robust value of 1. The rms noise in the final 5.5 GHz image was 7 $\mu$Jy bm$^{-1}$, and a radio continuum counterpart for \source{} was not detected, with a $3\sigma$ upper limit of $< 21 \mu$Jy.

\section{Results \& Analysis}

\subsection{Optical Spectra and Radial Velocities}
\label{sec:rv}

\begin{figure}
    \includegraphics[width=1.0\linewidth]{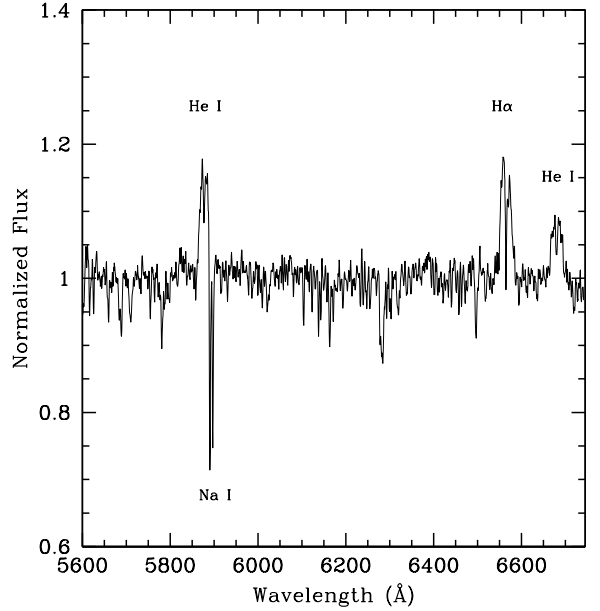}
    \caption{Continuum-normalized moderate-resolution SOAR/Goodman spectrum taken on
2017-11-15. Double-peaked emission lines of \ion{He}{I} at 5875 and 6678 \AA\ and \ion{H}{I} at 6563 \AA\ from the accretion disk are evident, as are the numerous narrow metal absorption lines from the secondary. The resonance Na I line is a blend of absorption from the secondary and the interstellar medium.}
    \label{fig:soarspec2}
\end{figure}

The SOAR spectra show the same qualitative features: a range of stellar photospheric absorption lines along with resolved, broad H$\alpha$ and double-peaked
He~I lines in emission that vary in prominence among the spectra
(Figure~\ref{fig:soarspec2}). 
We derived radial velocities of the secondary through a cross-correlation of the wavelength range 6050--6250 \AA, which is dominated by absorption lines. 
For the cross-correlation, we used an early-K star template, which on average provided higher cross-correlation peaks than mid-G or late-K templates.
There was no evidence for more than one set of absorption lines in any of the SOAR or MIKE spectra.
The emission lines associated with the accretion disk (and possibly partly
with the other structures in the system, such as the hot spot) form a second set of lines. 
While the double-peaked emission lines indicate the origin in an accretion disk around the white dwarf, it requires data with high 
signal-to-noise ratio in the line wings to make reliable radial velocity measurements of the white dwarf, 
free of interferences from other components such as the bright spot. Since this is not the case here, we effectively have 
a single-lined spectroscopic binary.

\begin{figure}
    \includegraphics[width=1.0\linewidth]{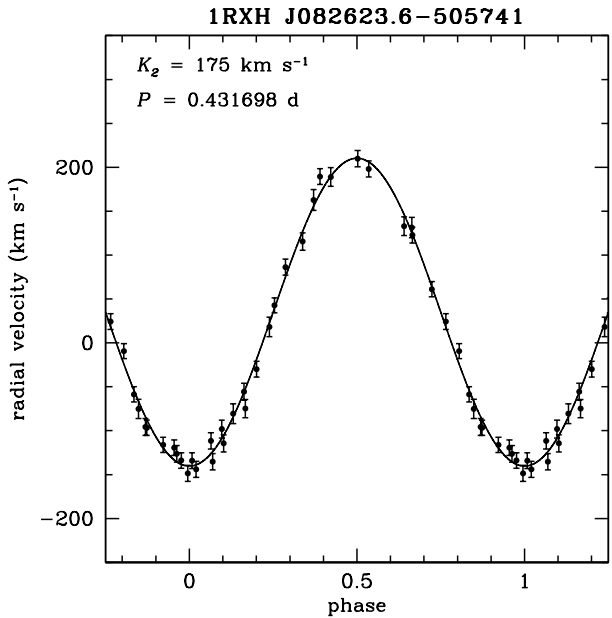}
    \caption{Radial velocity measurements of the secondary star, phased on the best-fit circular Keplerian model.} 
    \label{fig:opt_spectra}
\end{figure}

We first fit a circular model to the radial velocities using the custom Monte Carlo sampler \emph{TheJoker} \citep{2017ApJ...837...20P}. 
Each of the posterior distributions is close to normal, and the samples are essentially uncorrelated except for a weak correlation between the period $P$ and epoch of the ascending node of the white dwarf $T_0$. 
The best-fitting parameters are: $P = 0.4316979(17)$\,d, 
$T_{0~{\rm BJD(TDB)}} = 2457758.78073\pm0.00102$\,d, semi-amplitude $K_2 = 175.3\pm2.4$\,km\,s$^{-1}$, and systemic velocity $\gamma = 35.0\pm1.8$\,km\,s$^{-1}$.  This best-fitting circular model had 
$\chi^2/N_{\rm d.o.f.} = 35.7/30$ 
with an r.m.s. of 9.7\,km\,s$^{-1}$, and is shown in Figure~\ref{fig:opt_spectra}. This orbital period value matches that found from the optical photometry (Section \ref{sec:optobsphot}) within the relative precision of the two measurements.

An eccentric fit with six free parameters had a posterior eccentricity ($e$) distribution peaked at 0, with a median $e=0.014\pm0.013$ and a slightly worse 
$\chi^2/N_{\rm d.o.f.}=33.7/28$ 
than the circular fit. 
Hence, there is no compelling evidence for a deviation from the simple circular model. 

Fitting only the spectra obtained at orbital phases $\phi = 0.125$--0.375 (the side of the secondary without direct heating) gives $K_2 = 183\pm8$\,km\,s$^{-1}$, identical within the uncertainties to the value for the full fit.
We find no evidence that heating is meaningfully affecting our original $K_2$ measurement, which we use for the remainder of the paper.

For the subset of spectra with well-defined H$\alpha$ emission profiles, we find a mean FWHM of $1400\pm80$ km s$^{-1}$ 
and a mean peak separation of the blue and red H$\alpha$ peaks of $760\pm60$ km s$^{-1}$. 
These values are within the range of those observed for accreting white dwarfs (e.g., \citealt{2015ApJ...808...80C,2016ApJ...822...99C}). 
The mean equivalent width of H$\alpha$, $\sim 4-5$ \AA, is relatively low for an accreting white dwarf, 
but consistent with an evolved, relatively luminous secondary (see Section~\ref{sec:evolveddonor}).

\subsection{Rotational Velocity}
\label{sec:rotv}

While our typical SOAR spectra have a modest FWHM resolution of about 85\,km\,s$^{-1}$, in some spectra there is nevertheless evidence for line broadening, as would be present if the projected rotational velocity ($v \sin{i}$) of the secondary was being observed. 
Given that the synchronization timescale for \source{}~is $\lesssim10^{3}$ years at the orbital period and mass ratio derived above, we expect the visible secondary to be tidally locked.

To test if the absorption lines associated with the donor star are indeed broadened, 
we first analyze the single high-resolution MIKE spectrum of the binary (Section~\ref{sec:magellan}).
Following \cite{2017ApJ...851...31S}, 
we take a set of spectra of stars with similar spectral type and negligible rotation obtained with 
the same instrument setup. 
In order to simulate the rotational broadening of spectral lines, we convolve 
these template spectra with convolution kernels corresponding to 
a range of rotational velocities and assuming a standard limb darkening law. 
We then cross-correlate these convolved templates with the set of original unbroadened spectra 
to obtain a set of relationships ($N^2$ for $N$ templates) between the observed line broadening and $v \sin{i}$.
In effect, we construct the relationship of line FWHM vs $v \sin{i}$ from the templates
and then use it to derive $v \sin{i}$ from the observed FWHM.
One change we make, specific to this source, is to account for the motion of the secondary over 
the length of the exposure; while we deliberately obtained the spectrum relatively close to conjunction, 
the range of phases ($\phi = 0.407$ to 0.471) still corresponds to a motion of about 26\,km\,s$^{-1}$ 
using the best-fit orbital model described in Section \ref{sec:rv}. 
We simulate this by shifting the standard star spectra by small amounts appropriate to the varying velocity over 
the exposure and then averaging over these spectra. As it turns out this motion is much smaller than 
the measured value of $v \sin{i}$, and so this effect was essentially negligible.
The inferred mean intrinsic FWHM for the MIKE spectra is $133 \pm 12$\,km\,s$^{-1}$.

Using a range of orders with good signal-to-noise in the absorption features, but avoiding emission lines, 
telluric features, and spectral regions with substantial sky features, we found $v \sin{i} = 83.4\pm8.6$\,km\,s$^{-1}$. 
The uncertainty is dominated by the spread in values between different spectral orders rather than the spread among templates, 
which is much smaller. The precision of the measurement is somewhat lower than typical, which could be due to the modest 
signal-to-noise of the spectrum or to the presence of the accretion disk.

As a check on the result from this single MIKE spectrum, we repeated this $v \sin{i}$ analysis using the moderate resolution 
(1200\,l\,mm$^{-1}$) SOAR spectra, finding a value of $83 \pm 3$\,km\,s$^{-1}$.
The uncertainty is the error of the mean derived from 24 spectra with high enough signal-to-noise for this analysis 
(the measured standard deviation is $14$\,km\,s$^{-1}$).
That these values agree is reassuring, though it does not fully capture all systematic uncertainties in this analysis, 
such as the choice of limb-darkening law.

Given the short period and accreting nature of the system, it is reasonable to assume the non-degenerate secondary 
is indeed tidally synchronized so that its rotational period matches the orbital period. 
In this case, we can use the \citet{1983ApJ...268..368E} Roche Lobe approximation which relates 
$v \sin{i}$ and the measured radial velocity semi-amplitude $K_2$ to the mass ratio:
\begin{equation}
\label{eq:sam}
v \sin{i} = K_2 (1+q) \frac{0.49 q^{2/3} }{0.6 q^{2/3} + \textrm{ln}(1 + q^{1/3})}
\end{equation}
where $q = M_2/M_{1}$ is the mass ratio. 
Indeed, the observed axial rotation velocity of the secondary 
\begin{equation}
\label{eq:rotvel}
v \sin{i} = \frac{2 \pi R_2}{P} \sin{i}, 
\end{equation}
where $R_2$ is the secondary star radius. 
The observed maximum orbital velocity of the secondary 
\begin{equation}
\label{eq:orbvel}
K_2 = \frac{2 \pi a}{(1+q) P} \sin{i}, 
\end{equation}
where $a / (1+q)$ is the distance between the secondary star and the system center of mass, 
while $a$ is the binary separation. We obtain (\ref{eq:sam}) by dividing (\ref{eq:rotvel}) by (\ref{eq:orbvel})
and using the \citet{1983ApJ...268..368E} approximation for $R_2/a$.
The optically derived mass ratios using (\ref{eq:sam}), 
e.g., the ones obtained by \cite{2015ApJ...804L..12S} and \cite{2019ApJ...876....8S},
have been shown to be reliable for many compact binaries when compared with the direct mass ratio measurements from pulsar timing 
\citep{2019ApJ...872...42S}.
The values of $v \sin{i}$ and $K_2$ given above implies $q = 0.49\pm0.09$
for \source{}.

\subsection{Extinction}
\label{sec:extinct}

While the integrated line-of-sight reddening in this direction is high ($E(B-V)=0.82$; \citealt{SF11}), 
all the observational data are consistent with a lower extinction, likely due to the relatively nearby distance of the binary.
We estimate the extinction from the high-resolution MIKE spectrum, which allows a clean separation of intrinsic absorption lines from the interstellar ones. 
Using the \ion{Na}{I} calibration of \citet{2012MNRAS.426.1465P}, we found $E(B-V) = 0.10\pm0.04$. 
The calibration of \citet{1997A&A...318..269M}, which only uses the bluer stronger component of \ion{Na}{I}, 
gives a somewhat higher value of $E(B-V) = 0.18\pm0.05$. 
The results from light curve fitting favor a value at the lower end of this range, with $E(B-V) \lesssim 0.1$ (Section~\ref{sec:binarymodel}).

\subsection{Light Curve Modeling}
\label{sec:binarymodel}

The unabsorbed 0.2--10\,keV X-ray luminosity of $1.6 \times 10^{32}$ erg s$^{-1}$ in our XMM-Newton data, combined with previous Swift and ROSAT detections, shows a consistently high $L_X$ that implies \source{} is a compact binary. Further evidence is provided by luminous double-peaked \ion{He}{I} and \ion{H}{I} lines from optical spectra which are consistent with the presence of an accretion disk. 

Given the optical light curves show clear evidence for ellipsoidal variations (Figure~\ref{fig:opttimelc}), we 
undertook light curve modeling of the SMARTS $BVI$ data to constrain the binary inclination. We used the \textsc{Eclipsing Light Curve} \citep[\textsc{ELC};][]{Orosz00} code, 
which models filter-specific light curves in each band independently. In all our models we assumed a compact, 
invisible primary surrounded by an accretion disk, and a tidally distorted secondary in a circular orbit. 
The orbital properties were held to the values derived from our spectroscopic results. We assume the components are tidally locked.

For all models, we fit for the system inclination $i$ and the Roche Lobe filling factor and the intensity-weighted mean surface temperature of the secondary $T_{\rm cool}$. 
For the accretion disk, we fit for the inner disk temperature, 
the inner and outer radii of the disk, and the opening angle of the disk rim $\beta$. 
The accretion disk in the \textsc{ELC} code is assumed to be optically thick, 
an assumption that 
is in line with the current disk instability theory, but 
may not be entirely correct for a quiescent dwarf nova that has
emission-line dominated disk (see Section~\ref{sec:cvobs}).
The power-law index of the disk temperature profile was held to $-3/4$, 
a value appropriate for a ``steady-state'' disk \citep{1973A&A....24..337S,1981ARA&A..19..137P,1998ApJ...509..350W}, 
noting that disks in some quiescent dwarf novae are not in a steady
state and have flatter temperature profiles \citep[e.g.,][]{1989ApJ...341..974W,1996PASP..108...39O,2016MNRAS.463.3290R,2018AJ....156..153D}. 
We do not expect substantial irradiation of the secondary as it is already more luminous ($L_{\rm bol} \sim 3$--$4 \times 10^{33}$ erg s$^{-1}$) than the X-ray emission from the compact object/inner disk ($L_X/L_{\rm bol} \sim 0.05$), and indeed we see minimal evidence for irradiation in our light curve fitting.

A range of models can fit the data, and in particular the disk parameters are poorly constrained. Overall we find a best-fit inclination 
$i = 61.3^{\circ}$ $^{+6.8^{\circ}}_{-10.8^{\circ}}$  and Roche Lobe filling factor $0.90^{+0.08}_{-0.15}$ (corresponding to a secondary radius $R_2 = 0.78 R_{\odot}$). 
The secondary has an intensity-weighted mean $T_{\rm eff} = 5525^{+160}_{-167}$\,K, which varies from $\sim 4825$\,K on the gravity-darkened cool side facing the compact 
object to $\sim 5685$\,K on the warm side of the star facing away from the compact object. 
For the disk, we find inner and outer radii of 0.24 and 1.08 $R_{\odot}$ and inner and outer temperatures of 9720 and 3200\,K, 
along with a disk rim $\beta = 5^{+15^{\circ}}_{-3}$, but again these specific disk parameters should not be 
relied on as they are taken from one representative model and are not well constrained.
In particular, the relatively large inner disk radius should not be taken as evidence for a truncated 
disk, as this radius is not well constrained due to the relatively small contribution
of the disk to the total optical light output of the system. 
The peak separation of the emission line profiles 
(Section~\ref{sec:rv}) corresponds to the Keplerian velocity at $0.80 R_{\odot}$ for the
value of $i$ derived from this model and the corresponding $M_1$ (Section~\ref{sec:mass}). 
A model fit using the above mean parameters provides a good representation of the
lightcurve data ($\chi^2/N_{\rm d.o.f.} = 696/671$; 
see Figure~\ref{fig:optphaselc}), 
with a residual scatter in the $V$-band light curve of about 0.02\,mag,
consistent with the estimated photometric errors in this band (Section~\ref{sec:optobsphot}). 
In this model the disk contributes about 20\% of the light in $V$ 
(corresponding to the absolute disk magnitude of $\sim 6.8$).

Assuming a reddening at the lower end of the range favored by the optical spectroscopy ($E(B-V) = 0.07$), the implied photometric distance is $1.53\pm0.12$ kpc, in excellent agreement with the Gaia parallax distance. 
A higher reddening (e.g., $E(B-V) = 0.18$) would imply a nearer distance of $\sim 1.34$ kpc that is less consistent with the parallax distance.
Hence, a lower reddening is favored. Fixing the Roche Lobe filling factor to 1.0 does not change the inferred component masses, but slightly lowers the temperature of the secondary, implying a nearer distance and increasing tension with the parallax distance. This could be partially offset by a lower foreground reddening; there is no strong evidence that the secondary substantially underfills its Roche lobe.

\subsection{Masses}
\label{sec:mass}

The inferred values of $P$, $K_2$, and $q$ immediately give a minimum primary mass $M_1 \sin^3 i = 0.54\pm0.07 M_{\odot}$.

Using the inclination inferred from the light curve modeling, we find a primary mass $M_1 = 0.80^{+0.34}_{-0.17} M_{\odot}$ and a secondary mass $M_2 = 0.40^{+0.21}_{-0.13} M_{\odot}$.

\section{Discussion and Conclusions}
\label{sec:discussion}

\subsection{The physical nature of \source{}: a cataclysmic variable}

The mass constraints on the compact object in \source{} 
($M_1 = 0.80^{+0.34}_{-0.17} M_{\odot}$) immediately rule out a black hole, and strongly favor a white dwarf over a neutron star. The X-ray spectrum of the source is well-fit by 
an optically thin thermal plasma with a moderate X-ray luminosity, typical of white dwarfs (see the review of \citealt{2017PASP..129f2001M}), and arguing against an accreting neutron star, which typically at these luminosities show some combination of 
blackbody-like and power-law emission (e.g., \citealt{Campana98,Chakrabarty14}).
The lack of $\gamma$-ray (Section \ref{sec:intro}) and radio emission (Section \ref{sec:radioobs}), are also consistent with the white dwarf identification. For the remainder of the paper, we proceed on the basis that the binary contains an accreting white dwarf (cataclysmic variable).

\subsection{A non-magnetic cataclysmic variable}

The persistent presence of double-peaked emission lines in the optical spectra of
\source{}, revealing the presence of an accretion disk, immediately implies that the source cannot belong to the polar (AM\,Her) class. This is because the strong magnetic field of the white dwarf in such a class fully prevents the formation of an accretion disk.

However, its presence does not rule out the possibility of an intermediate polar, in which the weaker magnetic field of the white dwarf only partially truncates the inner accretion disk (e.g., \citealt{1994PASP..106..209P}).
Typically, these systems are confirmed as such when white dwarf spin periods are observed in the X-ray or optical at (usually) $\lesssim 1/10$ the orbital period. 
We see no evidence of any periodic signal in the expected range in the \emph{XMM} X-ray light curve which could be associated with the spin of the white dwarf, 
and likewise the optical emission only appears to vary on the orbital period. The spin signal can be muted if the inclination is near face-on, 
but the optical light curve fitting shows a typical intermediate inclination rather than a face-on one. 
High-ionization emission lines such as \ion{He}{II} 4686\,\AA\ have been used as empirical diagnostics of the presence of 
a magnetic white dwarf \citep{Silber92}. This line is not clearly present in our data, though this diagnostic is typically used 
in systems where the secondary makes a lesser contribution to the spectra than in this binary. 
Finally, \source{} has an X-ray luminosity that would be atypical for an intermediate polar \citep{2014MNRAS.442.2580P}, 
consistent with neither the bulk of the population (which has $L_X \gtrsim 10^{33}$ erg s$^{-1}$; \citealt{2003AdSpR..32.2067M,2016MNRAS.460..513T,2019ApJ...880..128L}) nor the
``low-luminosity'' intermediate polars with $L_X \sim 10^{31}$ erg s$^{-1}$ \citep{2014MNRAS.442.2580P,2020ApJ...898L..40L}, 
however DO\,Dra and V1025\,Cen have X-ray luminosities in this intermediate range \citep{2019MNRAS.482.3622S}.

In summary, while we cannot conclusively rule out the presence of a moderately magnetic white dwarf in \source{}, there is no evidence that supports this, so we proceed under the assumption that this is a non-magnetic system.

\subsection{The mass transfer rate and the state of the disk}
\label{sec:mdot}

If we make the simplified assumption that  half of the accretion luminosity is dissipated at the boundary layer and emerges in the form of X-rays, 
then for a $0.8 M_{\odot}$ white dwarf, the observed $L_X = 1.6 \times 10^{32}$ erg s$^{-1}$ implies 
$\dot{M} = $\accretionratemin $M_{\odot}$ yr$^{-1}$.
However, this does not account for the fact that some of the boundary layer emission could instead appear in the UV, 
so should be seen as closer to a lower limit. The other half of the accretion luminosity should be dissipated in the accretion disk. 
Our light curve model also produces an estimate of the bolometric luminosity of the disk, which is $\sim 1.3 \times 10^{33}$ erg $^{-1}$, 
though much of this luminosity comes out in the blue/UV where it is less well-constrained by our $BVI$ light curves. 
Taken at face value, this luminosity would imply 
$\dot{M} \sim $\accretionratemax $M_{\odot}$ yr$^{-1}$
substantially higher than the estimate from the X-ray luminosity.

Even at this higher estimate of $\dot{M}$, the system is still a factor of $\sim 100$ below the critical accretion rate that would be needed to keep 
the disk consistently in a stable hot ionized state at this orbital period (e.g., \citealt{2018A&A...617A..26D}). 
We expect that the disk \emph{should} be unstable and show occasional dwarf nova outbursts. 
Interestingly, in our long SMARTS time-series of optical observations (spanning 16 months), 
as well as in data from the All Sky Automated Survey for SuperNovae (ASAS-SN) that extend 
from 2016 Feb to 2021 Nov \citep{2014ApJ...788...48S,2017PASP..129j4502K}, 
there is no evidence for dwarf nova outbursts. 
While neither of these time series is continuous due to seasonal gaps, 
it suggests that dwarf nova events in this binary are either rare, of exceptionally low amplitude, or both.

If the disk is unstable to dwarf nova outbursts, the mass transfer rate
derived from X-ray (boundary layer) reflects the disk-to-white-dwarf transfer rate. 
This might be lower than the evolutionary important donor-star-to-disk mass transfer rate. 
The excess mass will accumulate in the disk to be dumped on the white dwarf during the
dwarf nova outburst.
The optical luminosity reflects the rate of mass transport within the disk, 
particularly the outer disk where the temperature is right to emit visible light.
It may be close or lower than the donor-star-to-disk transfer rate.

\subsection{An evolved donor}
\label{sec:evolveddonor}

A Roche Lobe-filling (or near-Roche Lobe filling) K-type donor in a 10.4-hr orbit must be evolved, 
as its density is too low to be a main sequence star (e.g.,
\citealt{1998A&A...339..518B}; an earlier spectral type donor could have been a main sequence star at this period). 
The measured parameters of the secondary in \source{} are also inconsistent with a normal main sequence star: 
the luminosity and mean spectral type of the star suggest a late G to early K star, 
which would be expected to have a mass in the range $\sim 0.7$--$0.9 M_{\odot}$, 
compared to the actual estimated mass of $0.40^{+0.21}_{-0.13} M_{\odot}$
(Section~\ref{sec:mass}). 

A subgiant donor expanding on a nuclear timescale would sustain a mass transfer rate of 
$\gtrsim 10^{-10} M_{\odot}$ \citep{1983ApJ...270..678W}.
This can be reconciled with the (in-disk) mass transfer rate of \accretionratemax $M_{\odot}$ yr$^{-1}$ inferred for \source{}
if the donor has a very low-mass core, as the radii and luminosities of low-mass red giants are virtually
uniquely determined by the masses of their degenerate cores, 
independent of the envelope mass.
The cataclysmic variable population synthesis models of \cite{2015ApJ...809...80G} 
predict a small fraction of systems with an accretion rate of a few $\times 10^{-10} M_{\odot}$
at an orbital period of ten hours, while the majority of the population has
higher accretion rates at this period (see also \citealt{2016ApJ...833...83K}).
However, observationally, low mass transfer rate systems with long orbital periods are not uncommon, 
which we discuss in more detail in the next subsection.

An alternative possibility is
that the current secondary in \source{} was previously more massive than the white dwarf 
and underwent an episode of thermal timescale mass transfer to the white dwarf 
before evolving to its current configuration \citep{2002ASPC..261..242S,2004ApJ...601.1058I}. 
The magnetic propeller system AE\,Aqr may have formed in this manner \citep{2002MNRAS.337.1105S}. 
The mass transfer rates of such donors would likely be governed by magnetic braking at their current periods 
and thus are hard to predict, but likely substantially lower than if the donor were evolving as a normal low-mass subgiant.

Finally, we considered the possibility that the donor star is not Roche lobe filling at all, but instead feeds the accretion disk via a wind,
as recently argued for the 8.7-d orbital period system 3XMM\,J174417.2$-$293944, which is thought to have a subgiant donor and a white dwarf accretor \citep{Shaw20}. However, the mass loss rate needed to power such a ``focused wind" would need to be higher than for Roche Lobe overflow (since the primary captures only a fraction of the wind), and the needed wind mass loss rate of $\gtrsim 10^{-9} M_{\odot}$ is implausibly
high for a K dwarf star, even if rapidly rotating (e.g., \citealt{2011ApJ...741...54C}).

\subsection{Context and Implications}

Only a modest number of non-magnetic cataclysmic variables with orbital periods comparable to that of \source{} are known, 
and those with few or no dwarf nova outbursts are even rarer. This can be readily attributed to selection effects: 
non-magnetic systems are usually not luminous hard X-ray sources that can be revealed by all-sky surveys with Swift/BAT or INTEGRAL/IBIS, 
and without regular dwarf nova outbursts, the systems cannot be identified by 
optical transient surveys such as ASAS-SN or 
the Zwicky Transient Factory (ZTF; \citealt{2019PASP..131a8002B}) either. Furthermore, the secondaries dominate the optical light, 
making color selection less effective and reducing the amplitude of dwarf nova outbursts
making them less easily detectable.
At the maximum of a dwarf nova outburst, the disk of \source{} may have $M_V > 3$ 
(including projection and limb darkening effect for high inclination systems; see
\citealt{2011MNRAS.411.2695P}), so the amplitude of its dwarf nova outbursts could well be $< 2$ mag.

The lack of dwarf nova outbursts in \source{} is likely a direct consequence of the low $\dot{M}$ 
(Section~\ref{sec:mdot}) 
and large disk (Section~\ref{sec:binarymodel}) in this binary, such that long intervals are necessary to build up enough mass in the disk to trigger an outburst. 
The well-studied cataclysmic variable DX\,And 
(Gaia~EDR3 distance $585 \pm 7$\,pc; \citealt{2021AJ....161..147B}), 
which has a similar orbital period and secondary \citep{1993MNRAS.260..803D,1997A&A...327.1107B}, 
has an estimated (outburst-cycle-averaged) mass transfer rate of $\sim 2 \times 10^{-9} M_{\odot}$ yr$^{-1}$ \citep{2018A&A...617A..26D}, 
an order of magnitude higher than the in-disk mass transfer rate 
we derived for \source{}, and DX\,And shows regular dwarf nova outbursts \citep{Weber62}.

\source{} is more akin to the nearby 11-hr orbital period cataclysmic variable 
EY\,Cyg ($629 \pm 7$\,pc; \citealt{2021AJ....161..147B})
having a late G/early-K type secondary \citep{2007A&A...462.1069E}, $L_X \sim 10^{32}$ erg s$^{-1}$ \citep{2020AdSpR..66.1139N}, 
and a dwarf nova recurrence time $> 5$ yr \citep{2002AIPC..637...72T}. 
Other potentially similar systems include CXOGBS\,J175553.2$-$281633 
($866 \pm 40$\,pc; \citealt{2021AJ....161..147B}), 
a 10.3-hr orbital period cataclysmic variable discovered via follow-up of a quiescent X-ray source 
in the Chandra Bulge Survey \citep{2021MNRAS.502...48G}, and 
KIC\,5608384 ($363 \pm 2$\,pc; \citealt{2021AJ....161..147B}), 
an 8.7-hr cataclysmic variable discovered in the Kepler field. CXOGBS\,J175553.2$-$281633 showed no dwarf nova outbursts 
in over 4 years of photometric observations and has a typical $L_X \sim 10^{31}$--$10^{32}$ erg\,s$^{-1}$ \citep{2021MNRAS.501.2790B}, while 
KIC\,5608384 had a single 5 day dwarf nova outburst over 4 years of data \citep{2019MNRAS.489.1023Y}. 
These systems deviate from the relation of \cite{2015MNRAS.448.3455B}
predicting that X-ray bright dwarf novae should have frequent outbursts.

It is notable that \source{}, CXOGBS\,J175553.2$-$281633, and KIC\,5608384 were discovered though systematic follow-up of all of 
the X-ray or optical sources in a specific field. Only EY\,Cyg was discovered via a dwarf nova outburst 
\citep{1928AN....233...33H}. \source{} is the most distant of the four similar systems. 
If a system like \source{} existed within 150\,pc, that would be a prominent ROSAT all-sky survey source 
and optical identification would have been easy. The fact that none is known (the nearest analog,
KIC\,5608384, is over 350\,pc away) while 42 typical cataclysmic variables are known within
150\,pc \citep{2020MNRAS.494.3799P} suggest that these long-period, low mass-transfer binaries have 
a much lower space density than typical cataclysmic variables.
At the same time, {\it it seems clear that long period, low mass-transfer binaries are underrepresented in 
the current cataclysmic variable samples due to selection effects in dwarf nova, color, or X-ray
catalogs.} More systematic searches are needed to quantify the degree of underrepresentation, 
and hence the likely space density of these systems. 

The low mass-transfer rates for these binaries are not well-explained by theory. 
For example, \citet{2019MNRAS.489.1023Y} construct custom MESA models of the evolution of KIC\,5608384, 
and find a predicted current mass-transfer rate that is a factor of 20 higher than the observed rate 
of $\sim 3 \times 10^{-10} M_{\odot}$ yr $^{-1}$. For any individual system, it is difficult to rule out stochastic variations 
around the mean mass-transfer rate, but the accumulating number of cataclysmic variables with these properties means 
that the observed mass-transfer rates should be taken more seriously. These values affect not only the interpretation of the current state 
of the binaries, but also  their future evolution, such as whether they become degenerate
AM\,CVn systems or not \citep[e.g.,][]{2015ApJ...809...80G,2016ApJ...833...83K,2019MNRAS.489.1023Y}. 

There is hope for improved discovery of long period, low mass-transfer cataclysmic variables. 
The all-sky soft X-ray survey of eROSITA should allow better identification of
non-magnetic systems compared to existing X-ray surveys 
(\citealt{2012MmSAI..83..844S}; for initial results on individual sources see
\citealt{2021arXiv210705611Z,2021arXiv210614538S,2021arXiv210614540S}), 
and the long time baselines of deep photometric surveys such as ZTF and the Rubin Observatory \citep{2021arXiv210801683B} 
will give increasing sensitivity to cataclysmic variables with rare dwarf nova outbursts.




\section*{Acknowledgements}
We are deeply grateful to Dr.~Kristen Dage and Dr.~Ryan Urquhart for their
advice on analyzing XMM-Newton data. We thank the referee for 
detailed comments that helped to improve this manuscript.

We acknowledge support from NSF grants AST-1714825 and AST-1751874 and the Packard Foundation.
KVS is supported by NASA Swift project 1821098 and XMM-Newton project 90327.
RLO acknowledges financial support from the Brazilian institution CNPq (PQ-312705/2020-4).
CH is supported by NSERC Discovery Grant RGPIN-2016-04602.

Portions of this work was performed while SJS held a NRC Research Associateship award at the Naval Research Laboratory. 
Work at the Naval Research Laboratory is supported by NASA DPR S-15633-Y.

This work is based on observations obtained with \textit{XMM-Newton} an ESA science mission 
with instruments and contributions directly funded by ESA Member States and NASA. 

Based on observations obtained at the Southern Astrophysical Research (SOAR) telescope, 
which is a joint project of the Minist\'{e}rio da Ci\^{e}ncia, Tecnologia, Inova\c{c}\~{o}es e Comunica\c{c}\~{o}es (MCTIC) do Brasil, 
the U.S. National Optical Astronomy Observatory (NOAO), the University of North Carolina at Chapel Hill (UNC), and Michigan State University (MSU).

The Australia Telescope Compact Array is part of the
Australia Telescope National Facility which is funded by the Australian Government for operation as a National Facility managed by CSIRO.

\facilities{XMM, Swift, ROSAT, SOAR, Magellan:Clay, CTIO:1.3m, ATCA}

\bibliographystyle{aasjournal}
\bibliography{main}

\appendix

\section{$H_m$ periodogram and power spectrum for the photon arrival times data}
\label{sec:hm}

The technique widely used to search for periodicities in X-ray and
$\gamma$-ray photon arrival times is often referred to as ``$H_m$-test''.
In fact, this is a periodogram closely related to the power spectrum with
the associated estimate of probability of finding a periodogram peak greater
than a specified value at a given trial period from randomly arriving
photons. Using photon arrival times (``event file'') for the period search 
has an advantage over the photon flux as a function of time measurements
(``lightcurve''). When the number of photons is low, one has to use wide
time bins (that would include many photons) to get an accurate measurement of
the photon flux. The width of the bin limits the time resolution and,
therefore, one's ability to find periods comparable or shorter than the bin
width. If photon arrival times are used directly, one's ability to find
short periods is fundamentally limited only by the detector's time
resolution. If the lightcurve is periodic, one can smooth the lightcurve 
in phase (summing up multiple periods) rather than in time.

\cite{1975Ap&SS..36..137D} defined a discrete Fourier transform of a function
as being equal to the convolution of the true Fourier transform of that function with 
a spectral window. The discrete Fourier transform is defined for an arbitrary 
data spacing that is encoded in the spectral window. 
The power, $P$, as a function of frequency, $\nu$, is defined as the squared modulus of the discrete Fourier
transform \citep[e.g.,][]{2014MNRAS.445..437M}:
\begin{equation}
\label{eq:power}
P(\nu) = \left [\sum_{i=1}^{N} f(t_i) \cos(2 \pi \nu t_i) \right]^2 + \left [\sum_{i=1}^{N} f(t_i) \sin(2 \pi \nu t_i) \right]^2
\end{equation}
The function, $f(t)$, is usually understood as photon (or energy) flux as a function of
time, $t$---a lightcurve. 
The transition from integration in the original Fourier transform definition 
to summation is possible as the function $f(t)$ is sampled at a set of $N$
moments in time, $t_i$ (where $i$ is the observation counter). 
This is equivalent to multiplying a continuous function $f(t)$ by 
a sum of Dirac $\delta$ functions $w_i = \delta(t-t_i)$ representing the window
function (the Fourier transform of the window function is the spectral
window), see equations (23) and (24) of \cite{1975Ap&SS..36..137D}.
Technically, nothing precludes $f(t)$ from
being either (a) 1 when a photon arrives 
and 0 at other times, or (b) constantly equal to 1 and having the window function encode the photon arrival times. 
The former is how \textsc{HEASoft} \textsc{XRONOS}
treats an event file, while the latter is how we think of
the event file in our interpretation: an irregularly sampled lightcurve
where most of the measurements are equal to 1 (the only exception is that
multiple photons may be registered with the same time stamp).

The case of $f(t)$ encoding the photon arrival times is also referred 
in the literature as the Rayleigh test \citep{1982Natur.296..833G,1983ApJ...272..256L}, 
which is usually explained in the following terms. 
For the Rayleigh test, each photon is represented by a unit vector with the
direction, corresponding to the photon arrival time phase for a trial period (trial frequency). 
If $\bar{R}$ is the sum of these unit vectors, the statistic $2 N \bar{R}^2$ is asymptotically 
distributed as $\chi^2_2$ for large $N$ \citep{mardia1972statistics,2000directional}, 
if the photons arrive at random phases of the trial period 
(allowing one to test observations against this null-hypothesis). 
Comparing this to (\ref{eq:power}), we see that for a given trial frequency $\nu$, 
$\bar{R}^2 = P(\nu)$ if $f(t_i)=1$ $\forall i$, while the factor $2 N$ 
provides a convenient normalization for the power to follow the $\chi^2$
distribution with 2 degrees of freedom. Now let us return for a moment from 
the photon arrival time to lightcurve space in order to illustrate 
how the definition of power (and the Rayleigh test) can me modified to account for 
a shape of the phased lightcurve.

Finding a power peak at frequency $\nu_{\rm peak}$ is equivalent 
to finding a least squares fit of a sinusoid (with angular frequency $2\pi\nu_{\rm peak}$) 
to the lightcurve (strictly speaking, this is true for the
modified relation proposed by \citealt{1982ApJ...263..835S}).
If the lightcurve is periodic with a single frequency, 
but the shape of the phased lightcurve differs from a sine wave,
analyzing the power spectrum (fitting a sine wave) would be a sub-optimal
strategy for finding $\nu_{\rm peak}$. A periodic lightcurve of a complex
shape can be approximated as a Fourier series \citep[e.g.,][]{2009A&A...507.1729D}, 
with only the first few Fourier components being sufficient for improving 
the efficiency of the period search. This is achieved by simply summing up 
the power at each trial frequency with that of its $m$ harmonics. 
This technique is known as ``$Z_m^2$ periodogram'' \citep{1983A&A...128..245B}, 
and it is widely used
\citep[e.g.,][]{2020ApJ...897...22P,2020A&A...643A..62D,2020arXiv201202280T}\footnote{A \textsc{Python} implementation of 
$Z_m^2$ periodogram by Yohan~Alexander may be found at \url{https://github.com/YohanAlexander/z2n-periodogram}.
We tested our code computing the discrete Fourier transform against this implementation.
An alternative implementation of $Z_m^2$ periodogram redefined for
binned pulsed profiles \citep{2021ApJ...909...33B} is available in \texttt{stingray} \citep{2019ApJ...881...39H}.}.
However, the phased lightcurve shape needed to choose an optimum value of $m$
is typically unknown in advance. The $H_m$ statistic is a heuristic method of
finding the optimal value of $m$ from the data by maximizing $Z_m^2$ value 
over $m$ for each trial frequency:
\begin{equation}
\label{eq:hm}
 H_m(\nu) = \max_{1 \le m \le m_\text{\rm max}} \left[c\times(1-m) + \sum\limits_{n = 1}^{m} P(n \nu)\right]
\end{equation}
where traditionally $c=4$, $m_\text{\rm max}=20$ \citep{2011ApJ...732...38K,2020ApJ...901..156N,2019ApJ...883...42N}
and $P$ is defined in (\ref{eq:power}).

First proposed by \citet{1989A&A...221..180D} 
(see also \citealt{2010A&A...517L...9D}) the $H_m$ periodogram has been employed to search for periods in high-energy
observations of cataclysmic variables \citep{2016ApJ...832...35L,2020A&A...637A..35S,2020A&A...639A..17W}, X-ray
\citep{2020ApJ...892....5K,2020MNRAS.498.4396H,2021MNRAS.500.1139H} and $\gamma$-ray pulsars
\citep{2017ApJ...846L..19P,2018SciA....4.7228C,2019ApJ...881...59B}, as 
well as arrival times of fast radio burst impulses \citep{2020Natur.582..351C}.
Despite the wide use, we could not find a suitable software implementation
of the technique, so we developed our own
\textsc{C} implementation as described below\footnote{\url{https://github.com/kirxkirx/patpc}}, following the description of
the original method by \cite{2011ApJ...732...38K}.
This code was also used by \cite{2021arXiv210803241S} to analyze XMM-Newton and NuSTAR data.

For a user-specified range of trial frequencies the code evaluates $H_m$
according to (\ref{eq:hm}). The trial frequencies are separated by 
$\Delta f = \Delta \phi / T$, where $T = t_N - t_0$ is the time span of the
observations and $\Delta \phi$ is the user-specified maximum allowed phase
shift between the first ($i=0$) and the last ($i=N$) observation when the
lightcurve is folded with the adjacent trial periods. The optimal value of 
$\Delta \phi$ depends on the lightcurve shape and signal-to-noise ratio: 
a small phase shift will be more noticeable if the lightcurve has sharp
features like a fast rise or a narrow eclipse and the amplitude of this
feature is well above the noise level. For many practical cases 
$\Delta \phi = 0.05$ is a good trade-off between the accuracy of period
recovery and the required computation time. Also, $\Delta \phi$ is the reciprocal
of the oversampling factor \citep[e.g.,][]{2006ApJ...653L..37P}, that is how
much finer is the frequency search grid compared to the ``natural''
resolution set by the Rayleigh criterion \citep[e.g.,][]{2003ASPC..292..383S}
which corresponds to the phase shift by one full period between the trial
periods ($\Delta \phi = 1$). The Rayleigh resolution criterion is also 
the conservative estimate of the period determination accuracy,
corresponding for the last observation in the lightcurve lag behind 
(or forging ahead) of the first observation by one full period. 

The concept of trial frequency spacing is related to the period determination accuracy.
The least detectable shift between the first and the last points of 
the lightcurve ($\Delta \phi$) limit the period determination accuracy. 
In other words, the two adjacent trial periods, $P$ and $P + \Delta \phi \times P$, 
that differ by less than $P_{\rm err} = \Delta \phi P^2 / T$ are indistinguishable.
Probing the whole range of trial periods (frequencies) at this high
resolution may be computationally expansive. Instead, one may sample the
range with the oversampling factor of a few (sufficient to find the
highest periodogram peak) and then repeat the search with the fine frequency
step corresponding to the minimum believable $\Delta \phi$ in a narrow range around the peak.
The problem with this approach to period accuracy estimation is that the
minimum detectable $\Delta \phi$ is hard to determine in practice, so 
the corresponding $P_{\rm err}$ will be an indication of the period
determination error, not a statistical one-sigma uncertainty. 

Alternatively, it should be possible to estimate the period uncertainty via bootstrapping. 
One could generate a set of mock data by randomly sampling real photon arrival times, 
so in each mock data set some of the photons from the original data will be
missing, while others will be duplicated. A scatter of periods derived from 
such mock data sets may be an indication of the uncertainty of the period 
derived from the original data. 

According to \cite{2010A&A...517L...9D}, for the standard values $c=4$, $m_\text{\rm max}=20$
in equation (\ref{eq:hm}) the probability of obtaining a value greater than $H_m$ by chance from
randomly arriving photons can be approximated as 
\begin{equation}
\label{eq:hmprob}
 p(>H_m) = e^{-0.4 H_m}.
\end{equation}
\cite{2011ApJ...732...38K} provides probability estimates for other values
of $c$ and $m_\text{\rm max}$. When estimating a significance of an $H_m$
periodogram peak, the probability estimate provided by equation (\ref{eq:hmprob}) 
has to be corrected for multiple trials. Indeed, we have multiple trial 
frequencies, however, the $H_m(\nu)$ (and $P(\nu)$) values measured at 
nearby frequencies are not independent. We follow the prescription of 
\cite{2003ASPC..292..383S} to estimate the number of independent trials as 
the smallest of 
(a) the number of observations $N$; 
(b) the number of frequencies in a search grid (if the trial period range is very narrow or $\Delta \phi > 1$); 
or 
(c) the number of frequencies in a grid spaced according to the Rayleigh resolution
criterion ($\Delta \phi = 1$; see Appendix~D of \citealt{1982ApJ...263..835S}).
We note that \cite{1986ApJ...302..757H} propose an alternative widely-quoted 
prescription for estimating the number of independent frequencies solely as a function of $N$
(see also the discussion by \citealt{2008MNRAS.388.1693F} and \citealt{2018ApJS..236...16V}).
While these authors discuss the number of independent frequencies in 
a context of lightcurves, the arguments should be equally applicable to
photon arrival times data. As we argue above, $f(t_i)$ in (\ref{eq:power})
may be considered a ``lightcurve'' while representing individual photon arrival
times: $f=1$ at times $t_i$ when the photons arrive and $f=0$ at all other times.
The discussion of the number of independent frequencies is concerned with the
number and distribution of $t_i$ and should not be affected by the units and
normalization of $f(t_i)$. 
The single-trial probability (\ref{eq:hmprob})
is multiplied by the number of independent frequencies (trials) 
to arrive at the final estimate of the chance occurrence probability of an $H_m$ periodogram peak.

One can circumvent the uncertain number of independent frequencies
problem using Monte-Carlo simulations \citep{2018JGRA..123.9204J}. 
The procedure would be to generate many sets of $N$ events with $t_i$ randomly 
distributed over the time interval(s) covered by the original observations. 
If the observations were continuous - that would be just the total duration of the
observations. If the observations were interrupted (for example by Earth
occultations or high particle background periods) the mock events should be
distributed only over the good time intervals when source photons could have been detected.
Running the desired period search algorithm (power spectrum -- Rayleigh test, $Z_m^2$ or $H_m$ periodogram) 
on the resulting mock event lists will allow one to judge how often a periodogram peak 
higher than the one found in the real data appears by chance in simulated random data.

In practice, the probability estimate (\ref{eq:hmprob}) corrected for
multiple trials, as well as the estimate derived from Monte-Carlo simulations,  
should still be treated as a lower limit. 
(This is analogous to the false alarm probability associated with the Lomb-Scargle
periodogram peak of \citealt{1982ApJ...263..835S}). If the source brightness
is changing non-periodically---for example, a secular brightening of a source---the photon arrival times will not be
random (as will be reflected by a low value of $p(>H_m)$). This problem is
known as the ``red-noise leakage'' \citep[e.g.,][]{2014MNRAS.445..437M}: 
a sidelobe of the spectral window may overlap with a spectral region that
has a lot of power. In the case of a long-term brightening of a source, 
this region would be at low frequencies. 
Chances of finding false peaks arising from the red-noise leakage are minimized if 
the trial periods a factor of 10--20 shorter than the
duration of observations are considered (so the sidelobes that may reach 
the low-frequency region are small). Another sign that a highly-significant
peak is produced by the red-noise leakage rather than actual periodic
variability is the presence of other peaks of comparable amplitude within
the factor of a few in frequency. One can check for the presence of
long-term variability by inspecting the binned lightcurve and if found,
expect the associated red-noise leakage problem. Finally, a periodicity
inherent to the observation technique (such as the orbital period of 
a satellite obtaining the data) may show up in the periodogram. The
knowledge of the observing technique and the periodicities inherent to it is
needed to distinguish them from a periodicity associated with the source.

In this appendix we described our implementation of the $H_m$ periodogram for 
photon arrival times data, highlighting its origin in \cite{1975Ap&SS..36..137D} 
discrete Fourier transform analysis of lightcurves (sets of timestamp---flux pairs). 
\cite{2021ApJ...909...33B} follow the argument in the opposite direction by  
generalizing the $H_m$ periodogram (originally defined for photon arrival
time data) to lightcurves. The relation between various statistics used for period 
search in photon arrival time and lightcurve data was also discussed by
\cite{2016ApJ...822...14B}. 

\end{document}